\documentclass{Interspeech}

\interspeechcameraready

\title{AdaKWS: Towards Robust Keyword Spotting with Test-Time Adaptation}

\author[affiliation={1,2}]{Yang}{Xiao}
\author[affiliation={3}]{Tianyi}{Peng}
\author[affiliation={4}]{Yanghao}{Zhou}
\author[affiliation={2}]{Rohan Kumar}{Das}

\affiliation{}{The University of Melbourne}{Australia}

\affiliation{}{Fortemedia Singapore}{Singapore}
\affiliation{}{Nanyang Technological University	}{Singapore}
\affiliation{}{National University of Singapore}{Singapore}

\email{yxiao9550@student.unimelb.edu.au, rohankd@fortemedia.com}

\usepackage{pifont}%

\newcommand{\cmark}{\ding{51}}%
\newcommand{\xmark}{\ding{55}}%
\usepackage{comment}
\usepackage{cite}
\usepackage{multirow}
\usepackage{balance}
\usepackage[table,xcdraw]{xcolor}
\begin{document}

\maketitle

\begin{abstract}
Spoken keyword spotting (KWS) aims to identify keywords in audio for wide applications, especially on edge devices. Current small-footprint KWS systems focus on efficient model designs. However, their inference performance can decline in unseen environments or noisy backgrounds. Test-time adaptation (TTA) helps models adapt to test samples without needing the original training data. In this study, we present AdaKWS, the first TTA method for robust KWS to the best of our knowledge. Specifically, 1) We initially optimize the model's confidence by selecting reliable samples based on prediction entropy minimization and adjusting the normalization statistics in each batch.
2) we introduce pseudo-keyword consistency (PKC) to identify critical, reliable features without overfitting to noise. Our experiments show that AdaKWS outperforms other methods across various conditions, including Gaussian noise and real-scenario noises. The code will be released in due course.
\end{abstract}
\vspace{2mm}

\section{Introduction}

Spoken keyword spotting (KWS)~\cite{lopezespejo2021deep} identifies specific keywords in audio input. It is essential in many popular applications, such as Apple Siri and Google Home, which are often used on edge devices~\cite{tabibian2020survey,smarthome,cl2}. Since KWS systems are always active, maintaining their performance with small-footprint models is vital for real-world use. Current deep learning-based KWS systems~\cite{bcresnet,kws1,kws2,kws3} achieve efficiency through smart design and are trained with a limited set of keywords to reduce computation and memory use while maintaining good performance. However, this strategy can cause the system to perform poorly when it encounters new or unfamiliar environments during actual use.

In many real-life applications, such as smart speakers, the performance of KWS systems often declines under low signal-to-noise ratio (SNR) and far-field conditions~\cite{wang2017trainable}. Although deep learning-based KWS models perform well when training and testing data are similar, room reverberation and various background noises challenge the models trained mainly on close-talking data due to limited real data collection. A traditional solution is to train models with pooled speech data from different environments~\cite{wu20j_interspeech,peng2024dark,jin24d_interspeech,lopez2021novel,lin2024advancing,AnalyticKWS}, but the labeled domain data is scarce and unsuitable after deployment. Another approach is to fine-tune the model. However, supervised fine-tuning is often impractical in real-world deployments because: 1) new environments may cause further domain shifts; 2) it also requires additional labeled data, which involves significant costs; and 3) users may prefer to keep their data private on local devices with limited computing power, making training-time fine-tuning infeasible. 

Due to limited domain-specific labeled data, methods like unsupervised domain adaptation (UDA)~\cite{ganin2015unsupervised} address the domain shift problem. Existing UDA approaches, such as domain adversarial training~\cite{domain1,domain2,hou2019domain}, knowledge distillation~\cite{lim2023joint,nguyen2021unsupervised,kim2021test}, and feature alignment~\cite{domain3}, effectively reduce domain shifts and improve KWS performance. However, these methods require access to source data and enough target domain examples for adaptation. This requirement creates the main limitation for real-world UDA applications: the source data may not be available during adaptation because of privacy and storage issues.

Recently, test-time adaptation (TTA)~\cite{tta,tent,sita,sar,eata} has gained attention because it can adjust models during prediction time using minimal target data without needing access to source data. A recent study, SUTA~\cite{suta}, explored TTA for automatic speech recognition (ASR) by minimizing prediction entropy, showing good results. However, applying TTA to KWS presents unique challenges due to two main factors. First, the small footprint of KWS models: KWS models are designed to be lightweight for efficient deployment on edge devices, meaning that even tiny changes can significantly impact their performance~\cite{zhang2018hello}. This sensitivity can lead to high entropy samples accumulating errors, especially when many existing ASR TTA methods~\cite{cea,sgem,lin2024continual} address domain mismatch to only a single type of noise. Second, the short and inconsistent test data: KWS systems operate on short utterances that lack extensive temporal information, making accurate predictions more difficult than ASR models. Therefore, it is vital to utilize samples that are less prone to incorrect predictions to reduce error accumulation.

To address these challenges, we propose AdaKWS, a test-time domain adaptation method tailored for robust KWS. Our approach tackles two main issues: 1) Small Footprint of KWS Models: We optimize model confidence by measuring prediction entropy and adjusting normalization statistics with channel-wise transformations, adapting only the sample with the small entropy to minimize accumulating errors. 2) Short and Non-discriminative Test Data: We introduce pseudo-keyword consistency (PKC), a metric that identifies harmful samples not detected by entropy. PKC measures how much the probability of a pseudo-keyword decreases after applying a transformation to the mel-frequency cepstral coefficients (MFCC), indicating the feature's impact on the model's prediction. We incorporate PKC by selecting and weighting samples more clearly rooted in keywords. This ensures that the model adapts to critical, reliable features (e.g., phoneme consistency) without overfitting to noise. Experiments on the Google Speech Command dataset demonstrate that AdaKWS improves robustness against Gaussian noise, and real-scenario noises showcasing its effectiveness across various acoustic domains. To the best of our knowledge, this is the first work that applies TTA to KWS highlighting contribution of the work. 


\begin{figure*}[!t]
\centering
\centerline{\includegraphics[width=\textwidth]{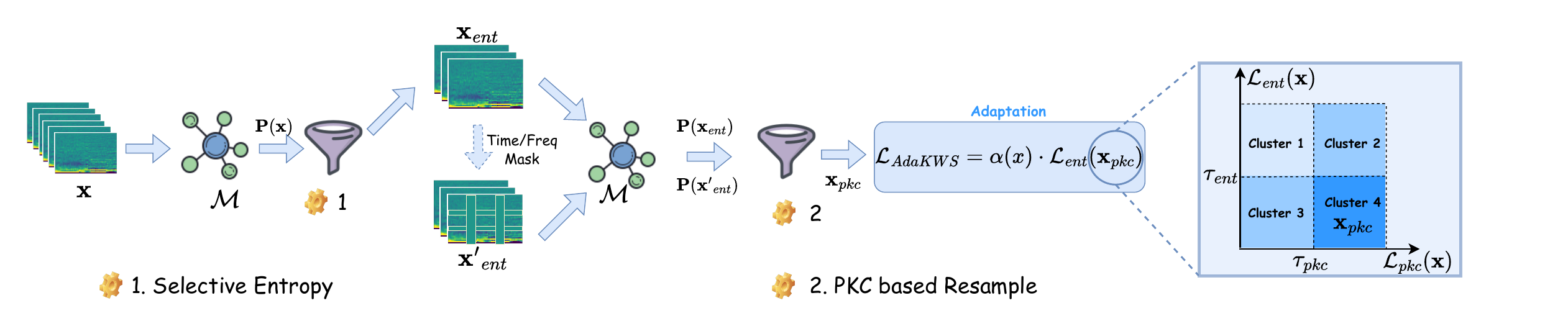}}
\vspace{-3mm}
\caption{Overview of the proposed AdaKWS method, including entropy- and PKC-based selection. The blue box highlights the subsets identified by these two approaches, and Cluster 4 corresponds to \(x_{pkc}\) in Eq. (6).}
\label{fig:overview}
\vspace{-3mm}
\end{figure*}

\section{Method}

\subsection{Problem Formulation}
Adaptation addresses generalization from train to test domain. We denote a trained KWS model as \(\mathcal{M}_\theta\), trained on source domain \(\mathcal{D}_{\text{train}}={(x^{\text{train}}_i, y^{\text{train}}_i)}^{N^{\text{train}}}_{i=1}\), and parameterized by the pre-trained weight \(\theta\), where \(x^{\text{train}}\in \mathcal{X}^{\text{train}}\) and \(y^{\text{train}}\in \mathcal{Y}\). The purpose of TTA is successfully adapting \(\mathcal{M}_\theta\) using the target domain test data \(\mathcal{D}_{\text{test}}={(x^{\text{test}}_i, y^{\text{test}}_i)}^{N^{\text{test}}}_{i=1}\), where \(x^{\text{test}}\in \mathcal{X}^{\text{test}}\) and \(y^{\text{test}}\in \mathcal{Y}\). Compared with fine-tuning and UDA, we cannot access the \(\mathcal{D}_{\text{train}}\) and \(y^{\text{test}}\) during the adaptation time. By not changing training, the TTA setting has the potential to require less data and computation for adaptation~\cite{tent}.

\subsection{Proposed AdaKWS Method}

\subsubsection{Selective Entropy Minimization}

Since test labels \(y^{\text{test}}\) are not available during the testing, we use an unsupervised, entropy-based loss function to adapt our model. Entropy minimization (EM) aims to sharpen the class distribution and is often applied in domain adaptation. Tent~\cite{tent} first introduces the idea of TTA by using EM and updating only the batch normalization (BN) layer’s parameters with batched inputs. In our proposed AdaKWS, we update BN parameters in the direction of minimizing \(\mathcal{L}_{\text{ent}}(x^{\text{test}})\) based on the entropy of \(x^{\text{test}}\):
\vspace{-4mm}
\begin{equation}
    \mathcal{L}_{\text{ent}}(x) = -\mathbf{P}(x)\cdot \log \mathbf{P}(x)=-\sum^C_{i=1} \mathbf{p}(x)_i \cdot \log 
 \mathbf{p}(x)_i
\end{equation} 
where \(\mathbf{P}(x)\) is the model’s output probability on \(x\) and \(C\) is the number of classes. For small-footprint KWS models, BN is often preferred over layer normalization used in ASR. BN statistics reflect a particular distribution, each distribution ideally needs its own statistics. However, estimating a single set of BN statistics from test mini-batches across multiple distributions can harm performance. Moreover, online Entropy Minimization can cause trivial collapsed solutions, in which the model predicts all samples as the same class. To address this, we pre-defined thresholds \(\tau_{\text{ent}}\) and based on the threshold we choose more reliable samples \(x_{\text{ent}}\) as:
\vspace{-1mm}
\begin{equation}
    x_{\text{ent}} = \{x|\mathcal{L}_{\text{ent}} < \tau_{\text{ent}}\}
\end{equation}

With this selective entropy minimization, the KWS models become more stable during the adaptation. 

\subsubsection{PKC based resample}

PKC is designed to address the limitations of entropy-based methods by evaluating the stability of model predictions under designed transformations. PKC identifies samples that rely on robust and consistent features, such as phoneme stability and spectral consistency, which are critical for KWS under noisy and unseen conditions.

In KWS systems, reliable feature extraction is essential for robust predictions, particularly in noisy environments. Entropy-based methods primarily focus on sharpening class probabilities but may fail to identify samples affected by transient noise or non-discriminative features. PKC complements entropy minimization by explicitly assessing the consistency of predictions when the input is perturbed by transformations in the time and frequency domains. In particular, time masking introduces temporal inconsistencies, emulating background interruptions or clipped speech. Frequency masking alters spectral representations, mimicking changes in audio equipment or environmental acoustics. 
Our method employs the following sample selection criteria:
\vspace{-3mm}
\begin{equation}
    \mathcal{L}_{\text{pkc}}(x,x') = {\mathbf{p}(x)_c - \mathbf{p}(x')_c}
\end{equation}
\begin{equation}
    x_{\text{pkc}} = \{x|\mathcal{L}_{\text{pkc}}(x,x') > \tau_{\text{pkc}}\}
\end{equation}
where \(\mathbf{p}(x)_c\) and \(\mathbf{p}(x')_c\) represent the model's confidence for pseudo-label \(c\) on the original input \(x\) and transformed input \(x'\), respectively. A smaller drop in prediction probability after transformation (i.e., low \(\mathcal{L}_{\text{pkc}}\)) indicates stability, and our selection focuses on these stable samples. Samples with \(\mathcal{L}_{\text{pkc}}(x,x') > \tau_{\text{pkc}}\) contribute positively to model updates, enhancing robustness and reducing the risk of overfitting to noise. PKC extends principles from spectral feature stability studies in speech recognition and domain adaptation~\cite{deyo}. Unlike traditional approaches that rely solely on a single prediction, PKC ensures that the model adapts to reliable, noise-robust features critical for KWS. Integrating PKC with entropy minimization enables AdaKWS to effectively identify and utilize robust samples, even in challenging acoustic environments.


\subsubsection{Overall Procedure of AdaKWS}
Our approach enables identified samples to make varying contributions to the model updates through sample weighting. Formally, we express the weighting function \(\alpha(x)\) as follows:
\vspace{-2mm}
\begin{equation}
    \alpha(x) = \frac{1}{\exp\{(\mathcal{L}_{\text{ent}}(x) - \sigma)\}} + \frac{1}{\exp\{-\mathcal{L}_{\text{pkc}}(x,x')\}}
\end{equation}

Where \(\sigma\) is a normalization factor. Our proposed AdaKWS method first performs sample selection by exploiting only the samples belonging to \(x_{\text{pkc}}\) and calculates the sample-wise weights \(\alpha(x)\) to prioritize samples. Then, the overall sample-weighted loss is given by:
\begin{equation}
    \mathcal{L}_{\text{AdaKWS}} = \alpha(x) \cdot \mathcal{L}_{\text{ent}}(x_{\text{pkc}})
\end{equation}

which combines entropy-based and PKC-based terms in both selection and weighting. 

\begin{figure}
    \centering
    \vspace{-2mm}
    \includegraphics[width=\linewidth]{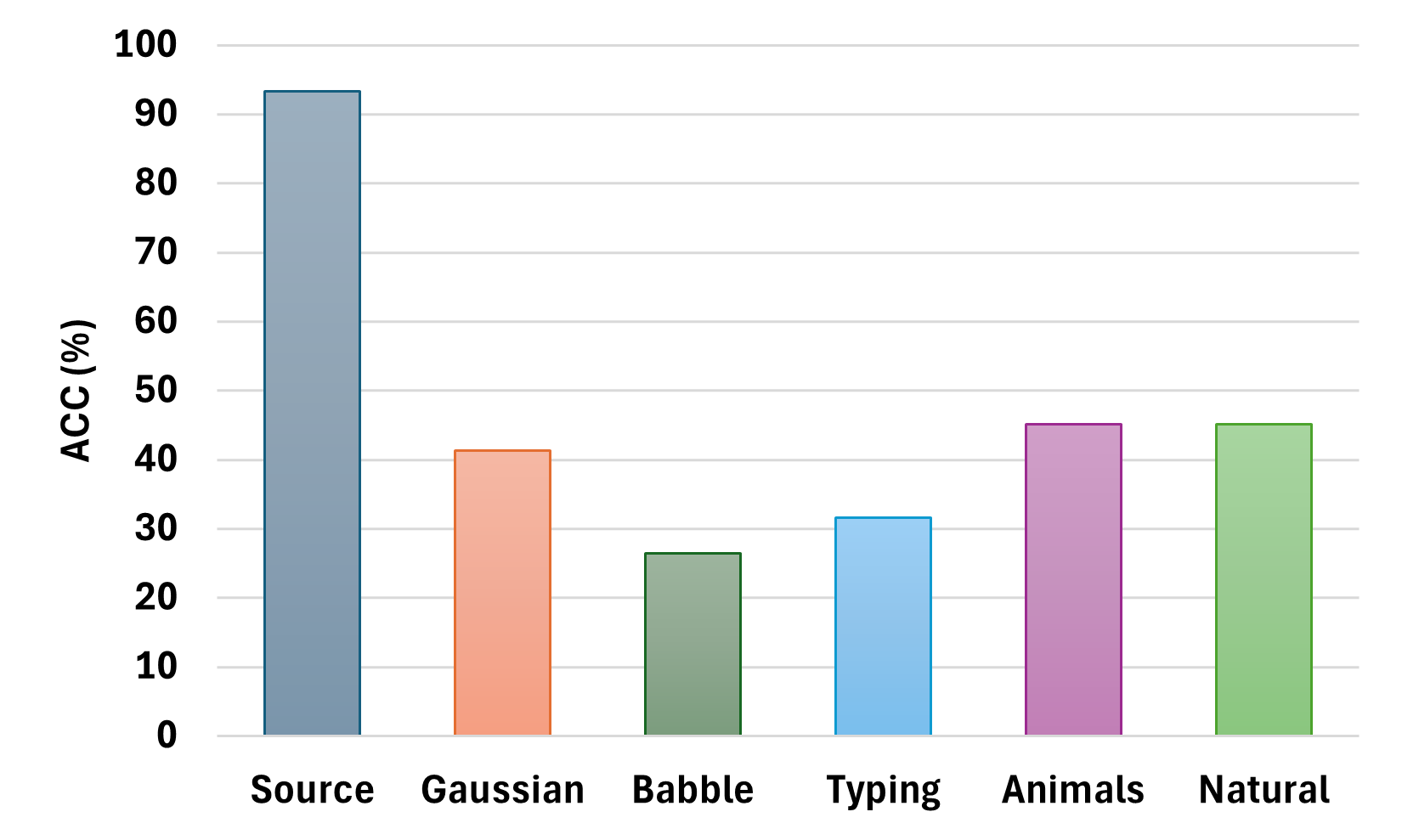}
    \vspace{-6mm}
    \caption{Robustness analysis of BC-ResNet-3 on GSC dataset under condition of (1) Gaussian noise (2) Babble and Typing from MS-SNSD dataset, (3) Animals and Natural from ESC-50 dataset. ``Source" shows performance on clean GSC dataset and ACC stands for accuracy.}
    \label{fig:enter-label}
\vspace{-4mm}
\end{figure}

\section{Experiments}
\subsection{Experimental Setup}
\textbf{Dataset: }We use the widely adopted Google Speech Commands~\cite{gsc} (GSC) dataset which includes 105,829 short audio clips, each lasting one second, covering 35 distinct keywords. Following established practices, we split each dataset into train (80\%), validation (10\%), and test (10\%) sets, with all audio sampled at 16 kHz. We evaluate our method under different testing conditions by adding two single-source noises and one multiple-source noise. First, we apply additive Gaussian noise at five severity levels, where higher levels mean stronger corruption. Next, we use real-world environmental noises (from the MS-SNSD \cite{snsd} test set) that include eight types such as Air Conditioner, Babble, and Vacuum Cleaner, each combined with the original audio at five signal-to-noise ratios (SNRs). Finally, for multiple-source noise, we use the ESC-50 \cite{esc} dataset, which contains 2,000 five-second recordings divided into 50 classes under five major categories, including Animals and Urban Noises. We broaden our investigation to realistic acoustic environments by introducing this more challenging domain.

\textbf{Baselines: }We measure the adaptation performance of our method by comparing it to several TTA baselines. First, test-time normalization (TBN)~\cite{bn1} updates batch normalization statistics with target data during testing. Next, Tent~\cite{tent} adapts the affine parameters of BN layers through entropy minimization. ETA~\cite{eata} filters out samples with high entropy to avoid noisy gradients, which can reduce performance. Finally, SAR~\cite{sar} employs sharpness-aware and reliable entropy minimization to handle practical issues like small batch sizes and online label shifts.

\textbf{Implementation details and metrics: }To evaluate our AdaKWS method, we use BC-ResNet-3~\cite{bcresnet} as the backbone, which is a lightweight CNN designed for on-device KWS. We extract 40-dimensional MFCC features as the input with 160 ms hop length. For our AdaKWS, we set the batch size to 128, and choose  \(\tau_{\text{ent}} = 0.4\), \(\sigma = 0.5\) and \(\tau_{\text{pkc}} = 0.05\). We apply two time-masks (max length 20) and two frequency-masks (max length 5) to transform the features for PKC calculation.

\section{Results}
\subsection{Impact of noisy domains}

We first investigate the impact of noise on small-footprint KWS models, focusing on BC-ResNet-3 trained on the GSC dataset. Robustness is analyzed under wild acoustic test conditions: (1) Gaussian noise (\(\delta = 0.03\)) added to the GSC test set, (2) Babble and Typing noises (-10dB) from the MS-SNSD dataset~\cite{snsd}, and (3) multiple noises like Animals and Natural sounds (-10dB) from the ESC-50 dataset~\cite{esc}. The ``Source" refers to the performance on the clean GSC test set without noise. While the model performs well on clean data, its accuracy (ACC) drops significantly under noisy conditions as Figure~\ref {fig:enter-label}. This highlights the need to adapt acoustic models for real-world deployment in noisy environments.

\subsection{Effect of adaptation with Gaussian noise}

\begin{table}[!t]
\centering
\caption{ Comparison of TTA methods under five severity levels (\(\delta\)) of Gaussian noise on the GSC dataset. Results are reported as accuracy (ACC, \%). The best results are \colorbox[HTML]{C4D5EB}{\textbf{highlighted}}.}
\vspace{-3mm}
\label{tab:noise1}
\resizebox{0.85\columnwidth}{!}{%
\begin{tabular}{lc|c|c|c}
\toprule
\multirow{2}{*}{\textbf{Methods}} & \multicolumn{4}{c}{\textbf{Gaussian noise \((\delta)\)}} \\ \cmidrule{2-5} 
                                  & 0.01     & 0.02     & 0.03 & Average    \\ \midrule
Unadapted                            &  70.96        & 53.46         &  41.34   & 55.92    \\
TBN                               &  83.96       &  75.10        & 67.37    &  75.48   \\
Tent                              & 84.44         &  75.70        & 68.59    &  76.24   \\
SAR                               & 83.63         & 74.59         & 66.99   &  75.07    \\
ETA                             &  83.69        &  74.97        & 67.36   &  75.34    \\ 
AdaKWS (Proposed)                 & \cellcolor[HTML]{C4D5EB}\textbf{84.66}         &  \cellcolor[HTML]{C4D5EB}\textbf{76.48}        & \cellcolor[HTML]{C4D5EB}\textbf{69.89}   &  \cellcolor[HTML]{C4D5EB}\textbf{77.01}    \\ \bottomrule
\end{tabular}%
}
\vspace{-3mm}
\end{table}

\begin{table}[!t]
\centering
\caption{Performance comparison of TTA methods under different MS-SNSD environmental noises (-10 dB). Results are reported as accuracy (ACC, \%).}
\label{tab:noise2}
\resizebox{\columnwidth}{!}{%
\begin{tabular}{lc|c|c|c}
\toprule
\multirow{2}{*}{\textbf{Methods}} & \multicolumn{4}{c}{\textbf{MS-SNSD Environmental Noise}} \\ \cmidrule{2-5} 
                                  & Typing     & Copy Machine     & Air Conditioner & Babble    \\ \midrule
Unadapted                            &    31.65      &  21.06        &  60.98   & 26.44    \\
TBN                              &   60.68       &  36.25        & 66.70    & 41.60    \\
Tent                              &   62.30        & 36.90          &  69.04   &  45.58   \\
SAR                               &   60.44       & 36.03         & 67.33   & 41.81     \\
ETA                              &   60.43       & 37.69         & 66.64   & 45.71     \\ 
AdaKWS (Proposed)                 &   \cellcolor[HTML]{C4D5EB}\textbf{62.33}       &  \cellcolor[HTML]{C4D5EB}\textbf{38.01}        & \cellcolor[HTML]{C4D5EB}\textbf{70.44}   &  \cellcolor[HTML]{C4D5EB}\textbf{49.56}    \\ \bottomrule
\end{tabular}%
}

\vspace{-4mm}
\end{table}

The results in Table~\ref{tab:noise1} demonstrate the effectiveness of the AdaKWS compared to baseline and existing TTA methods under different levels of Gaussian noise (\(\delta\)). The ``Unadapted" baseline shows the lowest average accuracy (55.92\%), indicating the unstable of KWS models in noisy environments. Among existing methods, Tent achieves the best performance (76.24\% average), followed closely by ETA (75.34\%) and SAR (75.07\%). AdaKWS, the proposed method, achieves the highest accuracy across all noise levels, with an average accuracy of 77.01\%. Notably, AdaKWS consistently outperforms others, especially at higher noise levels (e.g., \(\delta = 0.03\) ), showcasing its robustness in challenging conditions. These improvements are attributed to its novel use of PKC for identifying reliable features that enhance model adaptability and reduce overfitting to noise.
\begin{table*}[t]
\centering
\caption{Performance comparison of TTA methods under various noise categories from the ESC-50 dataset, including Animals, Natural, Urban, Human, and Domestic sounds. Results are reported for noise levels (-10 dB, 0 dB, 10 dB) and the average accuracy (ACC, \%).}
\label{tab:noise}
\resizebox{\textwidth}{!}{%
\begin{tabular}{lccc|ccc|ccc|ccc|ccc|ccc}
\toprule
\multirow{3}{*}{\textbf{Methods}} &
  \multicolumn{18}{c}{\textbf{ESC-50 Environmental Noise}} \\ \cmidrule{2-19} 
 &
  \multicolumn{3}{c|}{Animals} &
  \multicolumn{3}{c|}{Natural} &
  \multicolumn{3}{c|}{Urban} &
  \multicolumn{3}{c|}{Human} &
  \multicolumn{3}{c|}{Domestic} &
  \multicolumn{3}{c}{All}  \\ \cmidrule{2-19} 
 &
  -10 &
  0 &
  10 &
  -10 &
  0 &
  10 &
  -10 &
  0 &
  10 &
  -10 &
  0 &
  10 &
  -10 &
  0 &
  10 &
    -10 &
  0 &
  10 
   \\ \midrule
Unadapted & 45.13
   & 63.43
   & 74.76
   & 45.19
   & 61.75
   & 74.49
   & 48.76
   & 66.59
   & 77.56
   & 44.83
   & 62.32 
   & 74.29
   & 45.10
   & 61.25
   & 73.31
   & 45.74
   & 62.98
   & 74.82
   
   \\
TBN & 51.91
   & 67.83
   & 79.47
   & 57.22
   & 71.66 
   & 81.64
   & 59.99
   & 73.78
   & 83.26
   & 51.66
   & 68.35
   & 79.71
   & 55.76
   & 69.95
   & 79.78
   & 54.58
      & 69.31
   & 80.21
   
   \\
Tent & 52.25
   & 68.14
   & 79.66
   & 57.69
   & 71.99
   & 81.48
   & 60.47
   & 73.92
   & 83.25
   & 52.13
   & 68.71
   & 79.55
   & 56.71
   & 70.49
   & 79.76
   & 49.75
      & 68.15
   & 80.11
   
   \\
SAR & 51.55
   & 67.67
   & 79.09
   & 57.05
   & 71.14 
 & 81.34
   & 59.35
   & 73.54
   & 82.95
   & 51.76
   & 67.88
   & 79.27
   & 55.64
   & 69.72
   & 79.72
   & 54.12
         & 69.13
   & 79.99
   
   \\
ETA & 51.91
   & 67.32
   & 79.09
   & 56.87
   & 71.11 
   & 81.20
   & 59.75
   & 73.32
   & 83.09
   & 51.63
   & 67.90
   & 79.21
   & 55.56  
   & 69.62
   & 79.35 
   & 54.18
         & 69.28
   & 80.07
   
   \\ 
AdaKWS (Proposed) 
   & \cellcolor[HTML]{C4D5EB}\textbf{52.88}
   & \cellcolor[HTML]{C4D5EB}\textbf{68.76}
   & \cellcolor[HTML]{C4D5EB}\textbf{79.90}
   & \cellcolor[HTML]{C4D5EB}\textbf{58.27}
   & \cellcolor[HTML]{C4D5EB}\textbf{72.21}
   & \cellcolor[HTML]{C4D5EB}\textbf{81.96}
   & \cellcolor[HTML]{C4D5EB}\textbf{61.02}
   & \cellcolor[HTML]{C4D5EB}\textbf{74.86}
   & \cellcolor[HTML]{C4D5EB}\textbf{83.37}
   & \cellcolor[HTML]{C4D5EB}\textbf{52.22}
   & \cellcolor[HTML]{C4D5EB}\textbf{68.85}
   & \cellcolor[HTML]{C4D5EB}\textbf{79.84}
   & \cellcolor[HTML]{C4D5EB}\textbf{57.02}
   & \cellcolor[HTML]{C4D5EB}\textbf{70.86}
   & \cellcolor[HTML]{C4D5EB}\textbf{80.45}
   & \cellcolor[HTML]{C4D5EB}\textbf{54.96}
   & \cellcolor[HTML]{C4D5EB}\textbf{70.11}
   & \cellcolor[HTML]{C4D5EB}\textbf{81.07}
   
   \\ \bottomrule
\end{tabular}%
}
\vspace{-2mm}
\end{table*}
\subsection{Effect of adaptation with single environmental noise}
Table 2 evaluates the performance of the AdaKWS and other methods under various MS-SNSD environmental noises (-10 dB) using accuracy (ACC, \%). The ``Unadapted" model exhibits poor performance, particularly in challenging noise scenarios like ``Copy Machine" (21.06\%) and ``Babble" (26.44\%), emphasizing the need for robust adaptation. Among baseline TTA methods, Tent performs consistently well, with the highest accuracy in ``Air Conditioner" (69.04\%) and strong results in ``Babble" (45.58\%). However, AdaKWS outperforms all methods across most noise types, achieving notable improvements in ``Air Conditioner" (70.44\%) and ``Babble" (49.56\%), with competitive performance in ``Typing" (62.33\%) and ``Copy Machine" (38.01\%). These results highlight AdaKWS's ability to handle diverse and complex noise environments effectively.

\subsection{Effect of adaptation with multiple noises}

Table 3 presents a comprehensive comparison of TTA methods under diverse noise categories from the ESC-50 dataset, evaluating accuracy (ACC, \%) across three noise levels (-10 dB, 0 dB, 10 dB). ``All" presents using all five noise categories.  The ``Unadapted" method shows the weakest performance, with the lowest scores across categories such as ``Animals" (-10 dB: 45.13\%) and ``Domestic" (-10 dB: 45.10\%), highlighting its vulnerability to noise. Among the baselines, Tent consistently achieves high accuracy, excelling in ``Animals" (10 dB: 83.25\%) and ``Urban" (10 dB: 84.16\%) due to its effective entropy minimization for adapting to noise. SAR performs comparably well, particularly in ``Natural" (10 dB: 80.90\%), benefiting from sharpness-aware optimization. ETA offers competitive results in ``Human" (0 dB: 81.54\%), driven by its emphasis on stable adaptation. However, AdaKWS outperforms all methods overall, with the highest average accuracy across noise levels. Its consistent superiority across all categories and noise levels stems. Furthermore, AdaKWS is particularly strong in more complex categories like ``All" (0 dB: 70.11\%), where its adaptive mechanisms handle diverse acoustic characteristics effectively. This highlights AdaKWS as the most versatile and noise-resilient approach among all methods evaluated.

\subsection{Ablation study}

\begin{table}[t]
\centering
\caption{Ablation studies for the proposed AdaKWS method with different settings. All experiments use `Domestic' environments (-10dB) in ESC-50 noisy data.}
\vspace{-2mm}
\label{tab:ab-table}
\resizebox{0.9\linewidth}{!}{%
\begin{tabular}{ccc|c}
\toprule
 \textbf{Entropy Sampler} & \textbf{PKC Sampler} & \textbf{Rewighting}  & \textbf{ACC (\%)} \\ \midrule
 \cmark           &   \cmark          & \cmark           &              \cellcolor[HTML]{C4D5EB}{\bf 57.02}          \\ 
 \cmark           & \cmark           &      \xmark      &              56.84          \\ 
 \cmark           &     \xmark       & \cmark          &             55.94           \\ 
     \xmark      & \cmark           & \cmark                  &        56.59        \\  \bottomrule

\end{tabular}

}
\vspace{-6mm}
\end{table}

\begin{figure}[t]
    \centering
    \includegraphics[width=\columnwidth]{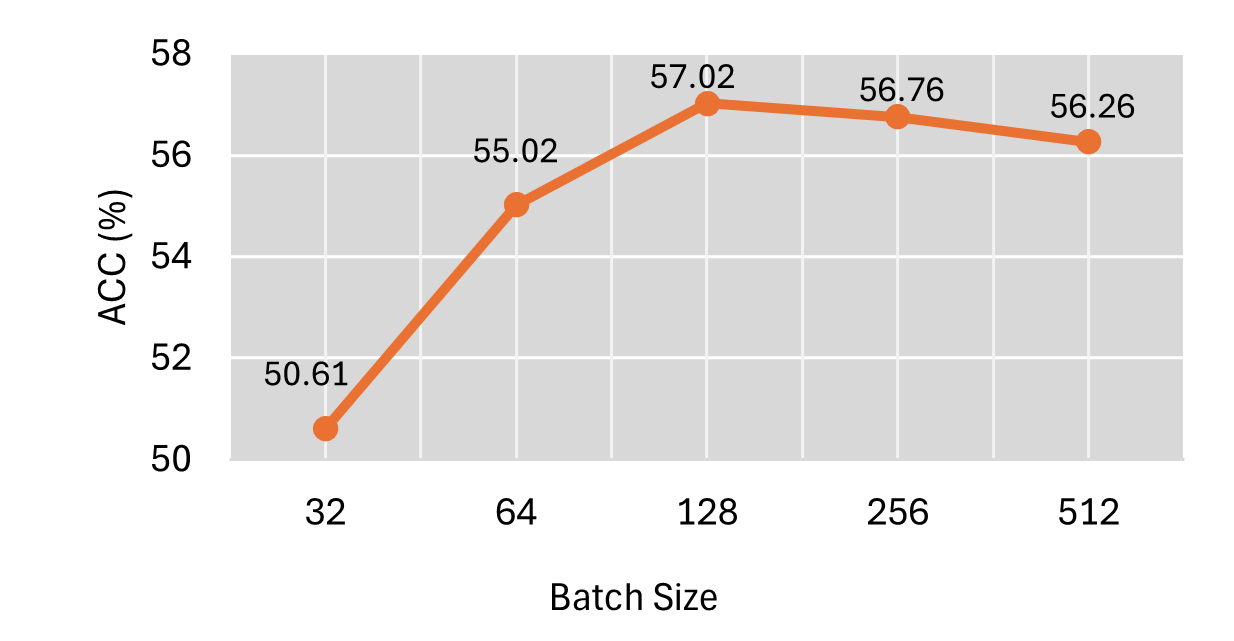}
    \vspace{-8mm}
    \caption{Comparable study across each test batch size for AdaKWS method. All experiments use `Domestic' environments (-10dB) in ESC-50 noisy data.}
    \label{fig:bs}
    \vspace{-7mm}
\end{figure}

Table 4 shows the results of the ablation study for AdaKWS in the ``Domestic" noise environment (-10 dB) from the ESC-50 dataset, isolating the contributions of its three key components: Entropy Sampler, PKC Sampler, and Reweighting. When all components are included, the proposed method achieves the highest accuracy (57.02\%), confirming the complementary roles of these mechanisms. Removing Reweighting results in a slight performance drop (56.84\%), suggesting its importance in balancing the influence of features during adaptation. Excluding the PKC Sampler further decreases accuracy to 55.94\%, highlighting its critical role in identifying reliable features. Without the Entropy Sampler, the accuracy remains relatively competitive (56.59\%), but the consistent decline underscores its contribution to sharpening class distributions. These results demonstrate that while each component independently improves performance, the integration of all three creates the effect that maximizes the robustness of AdaKWS in noisy environments.

Additionally, Figure 3 presents a comparative study of the AdaKWS method across different test batch sizes in the ``Domestic" noise environment (-10 dB) from the ESC-50 dataset. The results show a noticeable increase in accuracy as the batch size grows from 32 (50.61\%) to 128 (57.02\%), highlighting the advantage of larger batches for capturing robust feature patterns. However, beyond a batch size of 128, performance stops increasing, with slight decline observed at 256 (56.76\%) and 512 (56.26\%), suggesting diminishing returns for large batches. This trend indicates that while batch size positively impacts adaptation by improving statistical stability during channel-wise normalization, the large batches may introduce noise or reduce the method’s ability to focus on individual samples effectively. These findings emphasize the importance of selecting an optimal batch size (128 in this case) for balancing computational efficiency and performance in noisy environments.
\section{Conclusion}
 This study introduces AdaKWS, a novel TTA method for robust KWS in noisy environments. AdaKWS outperforms state-of-the-art TTA methods across complex noise conditions in various test domains. Key innovations include Selective Entropy Minimization based on channel-wise normalization and PKC for identifying critical and reliable features, which enhance adaptability and robustness, as confirmed by ablation studies. Our findings support research on entropy minimization and confidence optimization for adapting speech models during the test phase. Future work will explore cross-dataset generalization and resource efficiency. To the best of our knowledge, this is the first work that explores TTA to KWS.

\clearpage
\bibliographystyle{IEEEtran}
\bibliography{mybib}

\end{document}